\documentclass[prb,preprint,aps]{revtex4}
\usepackage{graphicx}
\usepackage{subfigure}
\usepackage{amsfonts}
\usepackage{amssymb}
\usepackage{amsmath}

\begin{document}
\title{Hyperspherical theory of anisotropic exciton }
\author{E. A. Muljarov,\cite{email} A. L. Yablonskii, and S. G. Tikhodeev}
\address{\\
General Physics Institute, RAS, Moscow 117942, Vavilov st., 38, Russia}
\author{A. E. Bulatov and Joseph L. Birman}
\address{\\
Physics Department, City College of New York, New York, NY 10031}
\date{\today}

\begin{abstract}
A new approach to the theory of anisotropic exciton based on Fock
transformation, i.e., on a stereographic projection of the momentum to the
unit 4-dimensional (4D) sphere, is developed.
Hyperspherical functions are used as a basis of the perturbation theory.
The binding energies, wave functions and oscillator strengths of elongated
as well as flattened excitons are obtained numerically.
It is shown that with an increase of the anisotropy
degree the oscillator strengths are markedly redistributed between
optically active and formerly inactive states, making the latter optically
active.
An approximate analytical solution of the anisotropic exciton
problem taking into account the angular momentum conserving terms is
obtained. This solution gives the binding energies of moderately anisotropic
exciton with a good accuracy and provides a useful qualitative description
of the energy level evolution.
\end{abstract}

\pacs{71.35.+z, 73.20.Dx}
\maketitle
\newpage
\relax

\section{Introduction}

The interest to the anisotropic exciton problem~\cite{Kitt54,Kohn55} has
been revived with the progress in the physics of semiconductor
heterostructures. In semiconductor superlattices the miniband formation
causes a strong mass anisotropy.~\cite{Bast88} In fact, the localization of
carriers inside quantum wells and their tunneling trough barriers can be
described in terms of anisotropic medium approximation as the effect of mass
renormalization. The dielectric constant becomes anisotropic also if the
superlattice constituent layers have different dielectric susceptibilities.
Recently such a formalism has been used in the theory of excitons in
short-period superlattices (see, e.g., Refs.~\onlinecite{Pere90,Ray93}).

The main complication of the uniaxial anisotropic exciton problem is that
the Coulomb potential symmetry is broken (the spherical symmetry as well as
the ``hidden'' one, the intrinsic property of the hydrogen-like system) so
that only the angular momentum projection and parity conserve. As a
consequence, the solution of the Schr\"odinger equation is no more
factorized into radial and angular parts and cannot be represented as a
finite combination of standard special functions.

The anisotropic exciton problem was first studied by Kohn and
Luttinger~\cite%
{Kohn55} (for donor states in silicon and germanium) by means of the
variational approach with allowance for a group symmetry of the particular
materials. Further theoretical studies~\cite%
{Hopf61,Whee62,Deve69,Sega67,Faul69,Akim67,Kane69,Bald70,Zimm71,Xia89,Depp92}
were focused on perturbative solutions of the anisotropic exciton problem.
For slightly anisotropic system Hopfield and Thomas~\cite{Hopf61} found the
first-order solution, treating the anisotropy of the kinetic energy as a
perturbation~\cite{foot1} linear in the anisotropy parameter. The effects in
a weak magnetic field also have been taken into account in this
approximation. For a moderate exciton anisotropy Wheeler and Dimmock~\cite%
{Whee62} used an expansion of the anisotropic potential over its asymmetric
part $z^2/r^2$ up to the second order in the anisotropy parameter terms,
thus calculating in part the second-order perturbation solution. This
partial diagonalization was completed by Deverin,~\cite{Deve69} who
considered the diagonal elements of the exact anisotropic kinetic energy
(for nondegenerate levels) as well as the transcendental solution of a
secular problem for degenerate levels. The full expansion of the anisotropic
potential was considered by Segal~\cite{Sega67}, where only the spherically
symmetric part of the full expansion was taken into account. Finally,
Faulkner~\cite{Faul69} performed calculations of donor energy levels by
means of Rayleigh-Ritz perturbation method containing numerous (depending on
hydrogen quantum numbers) variational parameters. Being included in the
radial part of hydrogen basis functions, these variational parameters served
as scaling factors depending on the anisotropy degree. In the limit of an
extreme anisotropy, the exciton binding energies were calculated~\cite%
{Kohn55,Akim67,Kane69} in adiabatic approximation. Following the method
suggested by Faulkner, Baldereschi and Diaz~\cite{Bald70} obtained similar
results and attempted to calculate excitonic oscillator strengths. The same
Rayleigh-Ritz method was used in Ref.~\onlinecite{Depp92} for calculations
of the energy levels of 2D anisotropic exciton.

Recently, an elegant model of fractional-dimensional space has been
developed [see Refs.~\onlinecite{He9091,Tang97} and references therein]. It
allows to treat self-consistently the bound as well as continuum states in
hydrogen problem of noninteger dimension. However, its direct applicability
to the anisotropic exciton problem is problematic. The reason is that the
fractional-dimensional hydrogen problem conserves the Coulomb degeneracy of
levels (so that the binding energies depend on the principal quantum number
only), whereas in reality the anisotropy lifts this degeneracy and restores
it only in 2D and 3D cases.

In spite of a long history of theoretical study, the investigation of the
optical properties of the anisotropic exciton is still not complete. For
example, the behavior of exciton oscillator strengths is very important for
the understanding the experimental absorption spectra. However, the
evolution of the oscillator strengths of the anisotropic exciton with the
increase of the anisotropy has not been investigated, for our knowledge,
with two exceptions: calculations for slightly anisotropic exciton\cite%
{Bald70} and simulations of optical spectra within an isotropic exciton
model.\cite{Pere95} One should note that none  of the approaches\cite%
{Bald70,Pere95} is able to describe the drastic changes of oscillator
strengths (due to the level anticrossings\cite{Faul69}) with increase of the
anisotropy reported in our paper.

In the present paper we develop\cite{our} a perturbation approach to the
uniaxial anisotropic exciton problem, based on the method of stereographic
projection of the momentum space to the unit 4D-sphere, proposed by Fock.~%
\cite{Fock35} We use the hyperspherical harmonics, i.e., the irreducible
representation of rotation group {\it O(4)} of a 4D-sphere, as a basis of
Brillouin-Wigner perturbation method.

This approach has a number of advantages and clarifies the physical
properties of the anisotropic exciton. (i) It allows us to utilize the
additional hidden symmetry of Coulomb potential for expansion of anisotropic
exciton wave function. Namely, for the bound exciton states the irreducible
representation of the full symmetry group {\it O(4)} constitutes a complete
set for such expansion. This expansion depends explicitly on the exciton
energy through scaling parameters which follow adiabatically the changes in
anisotropy. These parameters, similar to those introduced in the
Rayleigh-Ritz method~\cite{Faul69} (where they were defined by minimizing
the energy functional) are exactly determined in our method. As a result,
the hyperspherical functions turn out to be the most effective basis for
numerical calculations. (ii) Within Fock representation, the hydrogenic
spectrum with the level series limit transforms into an equidistant one,
which provides a good convergence of our method in a wide region of the
anisotropy parameter. (iii) The matrix elements of the perturbation are
found as analytical elementary expressions. (iv) This analytical form of
perturbation matrix elements allows us to construct a spherical
approximation with an analytical solution and to summarize exactly the rest
part of perturbation in the second order. This spherical approximation,
which works well in the region of a moderate anisotropy, turns out to be
very useful for qualitative classification of the energy levels.

We calculate numerically the energy spectrum, excitonic wavefunctions and
oscillator strengths for flattened as well as elongated excitons.

The paper is organized as follows. In Sec. II the expansion is formulated on
the basis of hyperspherical formalism and basic equations of the
perturbation method are derived. Results and discussions are presented in
Sec. III.

\section{Anisotropic exciton in Fock representation}

\subsection{Hyperspherical formalism}

The Hamiltonian of the uniaxial anisotropic exciton is given by
\begin{equation}
\label{Hamiltonian}\hat H=-\frac{\hbar ^2}{2\mu _{\perp }}\left( \frac{%
\partial ^2}{\partial x^2}+\frac{\partial ^2}{\partial y^2}\right) -\frac{%
\hbar ^2}{2\mu _{\parallel }}\frac{\partial ^2}{\partial z^2}-\frac{e^2}{%
\sqrt{\varepsilon _{\parallel }\varepsilon _{\perp }(x^2+y^2)+\varepsilon
_{\perp }^2z^2}}.
\end{equation}
Here $\mu $ is the reduced exciton mass, $\varepsilon $ is the semiconductor
dielectric constant, and subscripts $\parallel $ and $\perp $ refer to the
quantities along and normal to the axis of symmetry ($z$-axis),
respectively. In Eq.~(\ref{Hamiltonian}) both the kinetic and potential
energies are anisotropic. However, a dilatation $z\rightarrow z\sqrt{%
\varepsilon _{\parallel }/\varepsilon _{\perp }}$ makes the potential energy
spherically symmetric. In the effective atomic units
\begin{equation}
\label{units}{\rm Ry}^{*}=\frac{\mu _{\perp }e^4}{2\varepsilon _0^2\hbar
^2}%
,\ \ \ a_{{\rm B}}^{*}=\frac{\hbar ^2\varepsilon _0}{\mu _{\perp }e^2},
\end{equation}
where $\varepsilon _0=\sqrt{\varepsilon _{\perp }\varepsilon
_{\parallel }}$%
, Eq.~(\ref{Hamiltonian}) takes the form
\begin{equation}
\label{Schr-r}\left( \hat {{\bf p}}^2+\epsilon \hat p_z^2-\frac 2r\right)
\psi ({\bf r})=E\psi ({\bf r}).
\end{equation}
Here we introduced the perturbation parameter, $\epsilon =\gamma -1$,
connected to the anisotropy parameter,
\begin{equation}
\label{gamma}\gamma =\frac{\varepsilon _{\perp }\mu _{\perp }}{\varepsilon
_{\parallel }\mu _{\parallel }}
\end{equation}
($0<\gamma <1$ and $1<\gamma <\infty $ for, respectively, flattened and
elongated exciton), $\hat {{\bf p}}$ and $\hat p_z$ denote, respectively,
the dimensionless operators of momentum and its $z$-projection.

We investigate the bound states with eigenenergies $E_\nu <0$, measured in
$%
{\rm Ry}^{*}$, Eq.~(\ref{units}). It is convenient to introduce a parameter
(for each bound states $\nu $)
\begin{equation}
\label{p-nu}p_\nu =\sqrt{-E_\nu },
\end{equation}
which will play the role of the adiabatic parameter in the perturbation
theory. After the Fourier transform, Eq.~(\ref{Schr-r}) takes the integral
form
\begin{equation}
\label{Schr-p}(p^2+\epsilon p_z^2+p_\nu ^2)\psi _\nu ({\bf p})=\frac 1{2\pi
^2}\int \frac{\psi _\nu ({\bf p}^{\prime })}{|{\bf p}-{\bf p}^{\prime }|^2}%
d^3p^{\prime }.
\end{equation}
Following Fock's paper,~\cite{Fock35} we perform a stereographic projection
of 3D momentum space to the 4D unit sphere, ${\bf p}/p_\nu \rightarrow \vec
u
$, where the 4D vector $\vec u$ on the sphere is defined as
\begin{equation}
\label{u-vec}\vec u=\{{\bf u},u_n\}=\left\{ \frac{2p_\nu {\bf p}}{p^2+p_\nu
^2},\frac{p^2-p_\nu ^2}{p^2+p_\nu ^2}\right\} ,
\end{equation}
$p=|{\bf p}|$. In the hyperspherical coordinates, $(\alpha ,\theta
,\varphi )
$, the unit vector $\vec u$ takes the form
\begin{equation}
\label{stereo}\left\{
\begin{array}{l}
u_x=
\displaystyle{\frac{2p_\nu p_x}{p^2+p_\nu ^2}}=\sin \alpha \sin \theta \cos
\varphi , \\  \\
u_y=
\displaystyle{\frac{2p_\nu p_y}{p^2+p_\nu ^2}}=\sin \alpha \sin \theta \sin
\varphi , \\  \\
u_z=
\displaystyle{\frac{2p_\nu p_z}{p^2+p_\nu ^2}}=\sin \alpha \cos \theta , \\
\\
u_n=
\displaystyle{\frac{p^2-p_\nu ^2}{p^2+p_\nu ^2}}=\cos \alpha , \\
\end{array}
\right.
\end{equation}
and
\begin{equation}
\label{dif4}d^4\Omega =\sin ^2\alpha d\alpha \sin \theta d\theta d\varphi =%
\frac{8p_\nu ^3}{(p^2+p_\nu ^2)^3}d^3p.
\end{equation}
Let us introduce a new wave function
\begin{equation}
\label{Psi-connection}\Psi _\nu (\vec u)=\frac{(p^2+p_\nu ^2)^2}{4p_\nu
^{5/2}}\psi _\nu ({\bf p}),
\end{equation}
with normalization condition
\begin{equation}
\label{norm}\int |\psi _\nu ({\bf p})|^2d^3p=\int (1-\cos \alpha )|\Psi _\nu
(\vec u)|^2d^4\Omega =1.
\end{equation}
Then Eq.~(\ref{Schr-p}) takes the form
\begin{equation}
\label{Schr-u}\left( 1+\frac \epsilon 2\hat V\right) \Psi _\nu (\vec u)=%
\frac 1{p_\nu }\hat H_0\Psi _\nu (\vec u).
\end{equation}
Here $\hat H_0$ is the Hamiltonian of unperturbed (hydrogen-like) problem,
\begin{equation}
\hat H_0\Psi (\vec u)=\frac 1{2\pi ^2}\int \frac{\Psi (\vec u^{\prime })}{|%
\vec u-\vec u^{\prime }|^2}d^4\Omega ^{\prime },
\end{equation}
and $\hat V$ is the perturbation operator,
\begin{equation}
\label{V}\hat V=\frac{u_z^2}{1-u_n}=(1+\cos \alpha )\cos ^2\theta .
\end{equation}
If $\epsilon =0$ (or $\gamma =1$), Eq.~(\ref{Schr-u}) describes the
isotropic 3D exciton. As it was shown by Fock,~\cite{Fock35} the solutions
of the integral equation
\begin{equation}
\Psi ^{(0)}=\lambda ^{(0)}\hat H_0\Psi ^{(0)}
\end{equation}
are
\begin{equation}
\label{hyper}\Psi _{nlm}^{(0)}(\alpha ,\theta ,\varphi )=(-2i)^ll!\sqrt{%
\frac{2n(n-l-1)!}{\pi (n+l)!}}\sin ^l\alpha C_{n-l-1}^{l+1}(\cos \alpha
)Y_{lm}(\theta ,\varphi ),
\end{equation}
\begin{equation}
\label{p-n}\lambda _{nlm}^{(0)}=n,\ \ n=1,2,\dots ,\ \ l=0,\dots ,n-1,\ \
m=0,\pm 1,\dots ,\pm l.
\end{equation}
Here $C_k^m(x)$ are the Gegenbauer polynomials~\cite{Grad62} and $%
Y_{lm}(\theta ,\varphi )$ are the conventional spherical harmonics. The
hyperspherical functions, Eq.~(\ref{hyper}), afford the irreducible
representation of the full symmetry group {\it O(4)} of the hydrogen-like
system.~\cite{Band66} Due to the properties of irreducible representations,
the hyperspherical function are orthogonal and normalized as
\begin{equation}
\int |\Psi _{nlm}^{(0)}(\alpha ,\theta ,\varphi )|^2d^4\Omega =\int (1-\cos
\alpha )|\Psi _{nlm}^{(0)}(\alpha ,\theta ,\varphi )|^2d^4\Omega =1,
\end{equation}
in accordance~\cite{foot11} with Eq.~(\ref{norm}). It can be shown~\cite
{Podo29} that the standard hydrogen wave function $\phi _{nlm}^{(0)}({\bf
r})
$ with a given set of quantum numbers $(n,l,m)$ (see, e.g., in Ref.~%
\onlinecite{Land76}) can be Fourier transformed into the hyperspherical
function, Eq.~(\ref{hyper}).

\subsection{Formulation of Brillouin-Wigner perturbation theory}

We use the Brillouin-Wigner perturbation theory, i.e. the direct
diagonalization of a truncated Hamiltonian matrix in order to solve the
anisotropic exciton problem in the form of Eq.~(\ref{Schr-u}). The set of
the hydrogen bound states eigenfunctions is not complete and the scattering
states also must be taken into account. However, in Fock representation we
are able to construct a complete basis out of the set of the hydrogen bound
states. As it was shown in Ref.~\onlinecite{Band66}, the scattering states
are mapped on a two-sheeted hyperboloid in a 4D space with Minkowski
metrics, whereas the bound states are mapped into a unit sphere via the
transformation Eq.~(\ref{stereo}). Thus, the problems of the bound and
scattering states are mapped onto different subspaces, each of them to have
its own complete basis. The anisotropic problem is mapped into the same
subspaces through the transformation Eqs.~(\ref{u-vec})--(\ref
{Psi-connection}) for the bound states and the corresponding procedure (with
positive energies) for the scattering states. So, being interested in bound
states in the whole physical region $-1<\epsilon <\infty $, excluding the
points $\epsilon =-1$ (purely 2D exciton) and $\epsilon =\infty $ (purely 1D
exciton), we can use the hyperspherical harmonics Eq.~(\ref{hyper}) as a
complete set of basic functions.~\cite{foot00} As it immediately appears
from Eqs.~(\ref{Schr-u}) and (\ref{V}), the perturbation scheme converges
for $|\epsilon |<1$. For the opposite case of $\epsilon >1$ we can
reformulate the perturbation problem with the help of the transformation
\mbox{$p^2+\epsilon
p^2_z=(\epsilon+1)[p^2+(1/(\epsilon+1)-1)(p^2_x+p^2_y)]$.}
After this, we can redefine the effective atomic units Eq.~(\ref{units}) and
consider the operator \mbox{$(1/(\epsilon+1)-1)(p^2_x+p^2_y)$} as a
perturbation, thus providing the convergence for
\mbox{$|1/(\epsilon+1)-1|<1$.}

The eigenfunctions are expanded as
\begin{equation}
\label{Psi}\Psi_\nu(\alpha,\theta,\varphi)= S^{-1}_\nu \sum_s {\cal C}_s^\nu
\sqrt{n}\Psi^{(0)}_s(\alpha,\theta,\varphi), \ \ \ s=(n,l,m),
\end{equation}
where normalizing constants are defined as
\begin{equation}
\label{S-nu}S_\nu^2=\sum_{n,l} {\cal C}^\nu_{n,l,m}\left[n {\cal C}%
^\nu_{n,l,m} -\sqrt{(n+l+1)(n-l)}{\cal C}^\nu_{n+1,l,m} \right].
\end{equation}
Then, the Schr\"odinger equation takes the matrix form
\begin{equation}
\label{matrix-equ}\sum_{s^\prime}
\left(n\delta _{ss^{\prime}}+\frac{\epsilon}{2}{V}%
_{ss^{\prime}}\right) {\cal C}_{s^{\prime}}^\nu=\lambda_\nu{\cal C}_s^\nu,
\end{equation}
where
\begin{equation}
\label{lambda-E}\lambda_\nu=\frac{1}{p_\nu}=\frac{1}{\sqrt{-E_\nu}},
\end{equation}
and the perturbation matrix is
\begin{equation}
\label{Vss}V_{ss^{\prime}}=\sqrt{nn^{\prime}} \int\Psi^{(0)^\ast}_s(\alpha
,\theta,\varphi)(1+\cos\alpha )\cos^2\theta \Psi^{(0)}_{s^{\prime}}(\alpha
,\theta,\varphi) d^4\Omega.
\end{equation}
Nonvanishing matrix elements $V_{ss^{\prime}}$ are (see Appendix A)
\begin{equation}
\label{V-ll}{V}_{nn^{\prime}}^{ll;mm}={\cal Q}_{lm}\left\{n\delta
_{nn^{\prime}}+ \frac{1}{2}\sqrt{(n-l)(n+l+1)}\delta _{n+1\ n^{\prime}}+%
\frac{1}{2}\sqrt{(n-l-1)(n+l)} \delta _{n-1\ n^{\prime}}\right\}
\end{equation}
with
\begin{equation}
\label{B-lm}{\cal Q}_{lm}=\frac{1}{2}+\frac{1-4m^2}{2(2l-1)(2l+3)},
\end{equation}
and
\begin{equation}
\label{V-ll2}{V}_{nn^{\prime}}^{l\
l-2;mm}=\sqrt{\frac{[l^2-m^2][(l-1)^2-m^2]%
}{(2l+1)(2l-3)}} 2nn^{\prime}\sqrt{\frac{(n-l-1)!}{(n+l)!}\frac{%
(n^{\prime}+l-2)!}{(n^{\prime}-l+1)!}}{\cal F}_{nn^{\prime}}^l
\end{equation}
with
\begin{equation}
\label{Fnll1}\hspace{-2cm}
{\cal F}_{nn^{\prime}}^l
=\left\{
\begin{array}{l}
-1,\ \ \ \ n^{\prime}\leq n-2, \\
\\
\displaystyle \frac{n-l}{2n(2l-1)}\times\left\{
\begin{array}{ll}
\displaystyle\frac{n^2-4nl-l^2+n+3l-2}{2(n-1)}, & n^{\prime}=n-1, \\
&  \\
n-l+1, & n^{\prime}=n, \\
&  \\
\displaystyle\frac{(n-l+1)(n-l+2)}{2(n+1)}, & n^{\prime}=n+1, \\
&
\end{array}
\right. \\
\\
0,\ \ \ \ n^{\prime}\geq n+2. \\
\end{array}
\right.
\end{equation}
All the other matrix elements vanish.

The perturbation method in the form of Eq.~(\ref{matrix-equ}) is very
convenient. First of all, the perturbation $\epsilon \hat p_z^2$ is
invariant with respect to rotations around the $z$-axis and to the
transformation ${\bf p}\rightarrow -{\bf p}$. Thus, each perturbed state has
a definite parity and definite magnetic quantum number $m$, and the
perturbation problems Eq.~(\ref{matrix-equ}) can be solved separately for
different parity and $m$. It implies also that the summation over
$s^\prime $ in
Eq.~(\ref{matrix-equ}) and thereafter means that only the hydrogen states
with a given parity and magnetic quantum number have to be taken into
account. The time-conjugated states $\pm m$ are still degenerate. Secondly,
the precise form of perturbation matrix $V_{ss^{\prime }}$ provides more
rigorous selection rules. Namely, only the matrix elements with [see Eqs.~(%
\ref{V-ll})--(\ref{Fnll1})]
\begin{equation}
l^\prime=
\left\{
\begin{array}{ll}
l, & n^{\prime }=n,n\pm 1, \\
l-2, & n^{\prime }\leq n+1, \\
l+2, & n^{\prime }\geq n-1
\end{array}
\right.
\end{equation}
are nonvanishing.

The expansion (\ref{Psi}) corresponds to the following coordinate
representation of the anisotropic exciton wave function
\begin{equation}
\label{coordinate}\phi _\nu ({\bf r})=\frac{p_\nu ^{3/2}}{S_\nu }\sum_s{\cal %
C}_s^\nu n^2\phi _s^{(0)}({\bf r}p_\nu n),
\end{equation}
where $\phi _{nlm}^{(0)}({\bf r})$ are the standard hydrogen wave functions.
It follows from Eq.~(\ref{coordinate}) that the wave function of anisotropic
exciton takes the form of an infinite superposition of spherical harmonics
with radially dependent coefficients. The scaling factors $p_\nu $ in the
wave functions Eq.~(\ref{coordinate}), which are different for different
perturbed states and change adiabatically with $\epsilon $, play the role of
adiabatic scaling parameters in the perturbation theory. Moreover, the
coefficients $p_\nu $ are analogous to the parameters in the Rayleigh-Ritz
method. In contrast with previous works,~\cite{Faul69,Bald70} where $p_\nu $
have been calculated variationally, in our approach they are strictly
determined by Eq.~(\ref{lambda-E}). Finally, in spite of the energy scaling
factors in the basis functions, the effective Hamiltonian matrix in
Eq.~(\ref{matrix-equ}) is energy independent, thus allowing for the direct
diagonalization.

It can be seen from Eq.~(\ref{matrix-equ}) that in Fock representation the
spectrum of the unperturbed problem does not have a series limit. This fact
is favorable for the convergence of the perturbation theory. Moreover, the
spectrum is equidistant with respect to the hydrogen principal quantum
number $n$. The matrix Eq.~(\ref{V-ll}) is tridiagonal. The off-diagonal
matrix elements Eq.~(\ref{V-ll2}) with $n^{\prime}\sim n$ are rather
significant but do not exceed $n/2$, i.e., they are of the order of the
magnitude of the corresponding eigenvalues of the unperturbed problem. The
other nonzero elements, ${V}^{l\ l-2;mm}_{n n^{\prime}}\propto n^{-l}$,
decrease rapidly for fixed $n^{\prime}$ and $l$, and $n\gg n^{\prime}$,
$l>1$.
Thus, in numerical calculations we can take into account only the states
with lower $l$, and introduce a $\gamma$-dependent upper bound for the
orbital quantum number. Though the method provides a good convergence in a
large region of $\gamma$, it does not allow to avoid instabilities near $%
\gamma=0$ or $\gamma\rightarrow\infty$, where the perturbation scheme
becomes unstable, and a strong mixing of levels occurs.

We would like to emphasize that presented perturbation method can be easily
generalized for an arbitrary integer dimension $D\geq 2$ in accordance with
Ref.~\onlinecite{Band66}, where the method of stereographic projection has
been expanded to higher dimensions. In particular, for $D=2$ the standard
spherical harmonics $Y_{lm}(\theta ,\varphi )$ have to be used as a basis
and the operator $(\epsilon /2)\hat V=(\epsilon /2)(1+\cos \theta )\cos
^2\varphi $ --- as a perturbation. Here $\cos \theta =(p^2-p_\nu
^2)/(p^2+p_\nu ^2)$, $\tan \varphi =p_y/p_x$.

The problem of anisotropic exciton scattering states can be approached
analogously using hyperspherical harmonics on a two-sheeted 4D-hyperboloid
as a basis for the perturbation problem. The eigenvalues should be defined
with positive energies, instead of Eq.~(\ref{lambda-E}). However, the
eigenvalue problem [analogous to Eq.~(\ref{matrix-equ})] becomes more
complicated: we have to solve now a system of integral equations, because of
dependence on continuum quantum numbers.

One should note that the method of stereographic projection can be formally
generalized for the fractional-dimensional exciton problem, the exciton
binding energies coinciding with those obtained in Ref.~\onlinecite{He9091}.
However, due to the generalized hyperspherical symmetry conservation (the
anisotropy parameter now appears in a role of the fractional
dimensionality), the energy levels are Coulomb degenerate, as it was
mentioned above.

\section{Results and discussions}

Due to the symmetry properties of uniaxial anisotropic exciton Hamiltonian,
matrices with even and odd $l$ as well as with different $m$ can be
diagonalized independently. In contrast to the variational technique which
provides only the upper bound of the binding energies, the Brillouin-Wigner
perturbation method allows us to reach necessary precision by choosing a
sufficiently large matrix to be diagonalized. We perform our calculation
with a relative energy precision of $10^{-4}$. In order to provide this
precision in the calculation of the ground state energy for $0.6\leq \gamma
^{1/3}\leq 2$, hydrogen states with the principal quantum number up to 15
and orbital quantum number up to 6 must be taken into account. The numerical
procedure becomes unstable for $\gamma \rightarrow 0$ and $\gamma
\rightarrow \infty $. This non-convergency is caused by the fact that these
points, where the symmetry changes (to 2D and 1D, respectively), are
peculiar for the perturbation theory. The dimension change causes the
levels' degeneration, when a very large (divergent) number of levels is
mixed due to perturbation, and has to be taken into account. To calculate
the ground state exciton energy within a relative accuracy of $10^{-4}$, the
levels with principle quantum number $n\leq N$ should be taken into account.
In Fig.\thinspace 1 we show the numerically found dependence of $N$ on the
anisotropy parameter $\gamma $ (for $\gamma \leq 1$), which is approximately
logarithmic.

\begin{figure}
\vskip-1cm
\includegraphics[angle=0,width=0.5\linewidth]{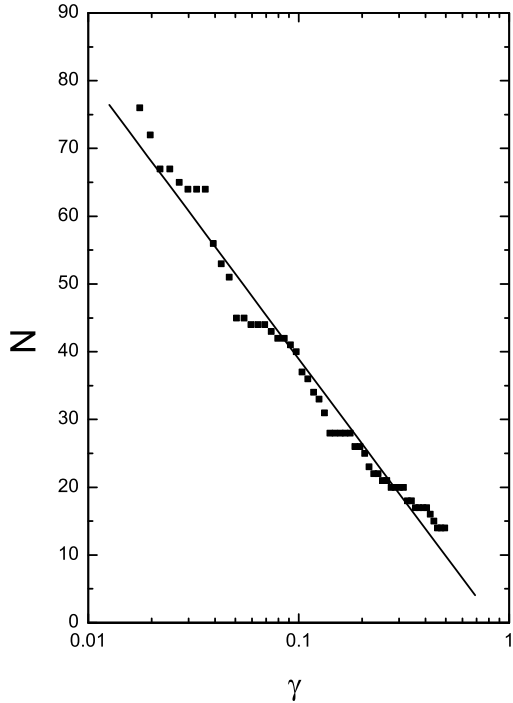}
\vskip-1cm
\caption{The maximum principle quantum number
$N$ of the states  used in numerical calculations
of the ground state exciton energy within a relative margin of
$\protect 10^{-4}$
as a function of the anisotropy parameter $\gamma $ ($\gamma \leq 1$).
Solid line shows the logarithmic approximation for $N$.}
\end{figure}
Note that at a rather strong anisotropy, when $\gamma \ll 1$, the ground
state exciton behaves as $E_0\approx -4+10.3\gamma ^{1/3}$, $|\phi
_0(0)|^2\propto \gamma ^{-1/3}$ and $\sqrt{\langle z^2\rangle }\propto
\gamma ^{-1/3}$, (see in Ref.~\onlinecite{Kohn55}). Thus, it is useful to
plot physical values in dependence on $\gamma ^{1/3}$ instead of $\gamma $.

\subsection{Energy levels}

\label{energies}

Figures 2 and 3 show the calculated eigenvalues $\lambda_\nu$ of Eq.~(\ref
{matrix-equ}), related to the exciton energies, $E_\nu=-1/\lambda_\nu^2$, as
functions of $\gamma^{1/3}$ for $\gamma\leq1$ (left panels); $%
\gamma^{-1/2}\lambda_\nu$ are shown as functions of $\gamma^{-1/3}$ for $%
\gamma\geq1$ (right panels). The multiplier $\gamma ^{-1/2}$ in the latter
case makes the effective Rydberg finite when $\mu_\perp\rightarrow\infty$.
The binding energies of $m=0$ even parity states and $m=1$ odd parity states
are shown, respectively, in Figs.\,2 and 3. Starting at $\gamma=1$ from $%
\lambda_\nu=\lambda^{(0)}_{nml}=n$, all the eigenvalues with the same $m$
and parity do not intersect when $\gamma$ changes (multiple anticrossings
occur due to the interaction between states) and approach the ground state
eigenvalue of 1D exciton~\cite{foot2} $\gamma^{-1/2}\lambda_\nu\rightarrow%
\gamma^{-1/2}\lambda^{{\rm 1D}}_0\rightarrow0$ (Figs.\,2 and 3, right
panels), when $\gamma\rightarrow\infty$. In the opposite case of $%
\gamma\rightarrow0$ all shown eigenvalues approach the ground state
eigenvalue $\lambda^{{\rm 2D}}_0=1/2$ of 2D exciton ($m=0$, Fig.\,2, left
panel) or the first excited state eigenvalue $\lambda^{{\rm 2D}}_1=3/2$ ($%
m=1 $, Fig.\,3, left panel). As it is clear from Fig.\,2, the ground state
eigenvalue dependence is almost linear over $\gamma ^{1/3}$ for $\gamma \leq
1$, and
\begin{equation}
E_0\approx-\frac{4}{(1+\gamma ^{1/3})^2}.
\end{equation}
The ground state which lies much lower than the excited states almost does
not interact with the latter. However, for the first excited state this
interaction becomes much more significant, and its energy dependence upon $%
\gamma ^{1/3}$ deviates from the linear one (cf. with dashed line in
Fig.\,2).
\begin{figure}
\vskip-8cm
\includegraphics[angle=0,width=0.99\linewidth]{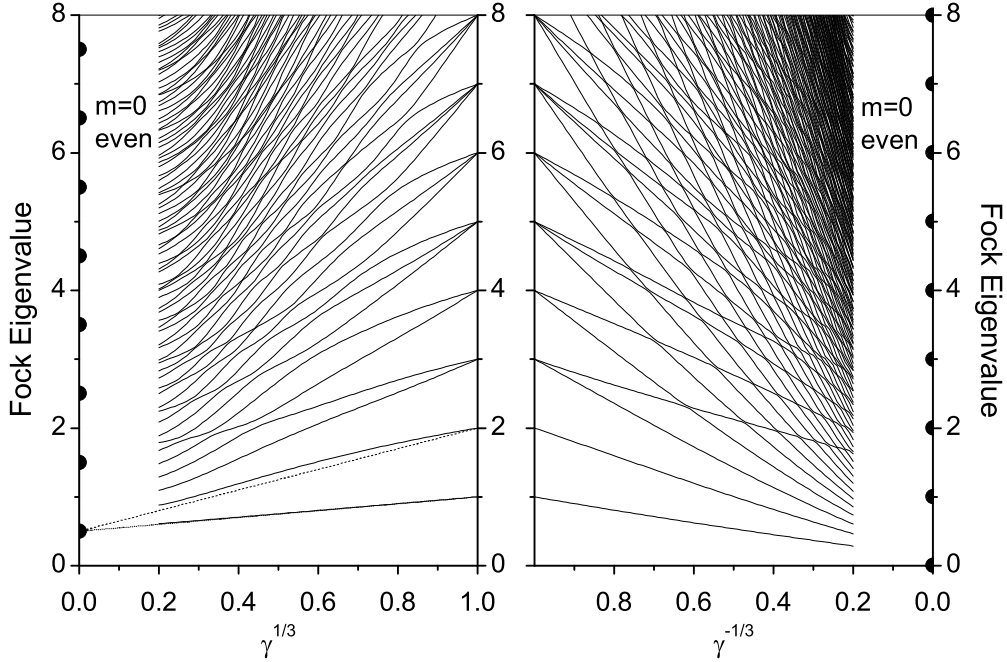}
\vskip-7cm
\caption{
Fock eigenvalues $\protect\lambda_\nu$ of $m=0$ even parity
states as functions of the anisotropy parameter
$\protect\gamma^{1/3}$, $\protect\gamma\leq1$ (left panel), and
$\protect\gamma^{-1/2}\protect\lambda_\nu$ as functions of
$\protect\gamma^{-1/3}$, $\gamma\geq 1$ (right panel).
Solid curves never intersect each other due to a small
anticrossing between the levels. The eigenvalues of purely 2D
exciton (left panel) and 1D exciton (right panel) are shown by
semicircles.
A linear
approximation of the ground and first excited state eigenvalues
is plotted by dotted and dashed lines, respectively.
}
\end{figure}
\begin{figure}
\vskip-8cm
\includegraphics[angle=0,width=0.99\linewidth]{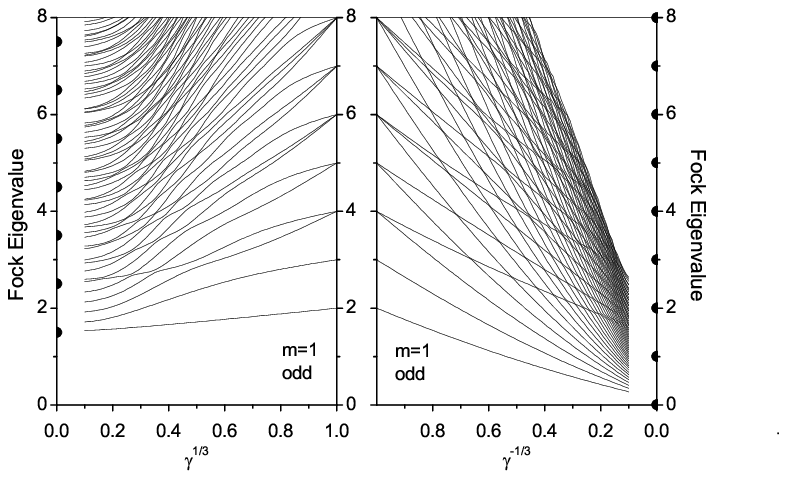}
\vskip-7cm
\caption{
Fock eigenvalues $\protect\lambda_\nu$ of $m=1$ odd parity
states as functions of the anisotropy parameter
$\protect\gamma^{1/3}$, $\protect\gamma\leq1$ (left panel), and
$\protect\gamma^{-1/2}\lambda_\nu$ as functions of
$\protect\gamma^{-1/3}$, $\protect\gamma\geq1$ (right panel).
}
\end{figure}
The ratio of the energy separation between the ground state and the first
excited state to the exciton binding energy is shown in Fig.\,4. Starting
from 3/4 for 3D-isotropic exciton $(E_{1S}-E_{2S})/E_{1S}$ decreases
monotonously with change of $\gamma $ and vanishes when $\gamma \to 0$ or $%
\gamma \to\infty$. Thus, this quantity can be considered as a measure of the
anisotropy of a system. Note that within the fractional dimensional model $%
(E_{1S}-E_{2S})/E_{1S}$ grows up as $1-[(D-1)/(D+1)]^2$ ($D$ is the
dimensionality). Thus, in the anisotropic model the transition from 3D
exciton to 2D or to 1D exciton differs completely from that of a system, in
which the carriers localization in one or two dimensions becomes stronger
and using of the fractional dimensional model is justified.

\begin{figure}
\vskip-1cm
\includegraphics[angle=0,width=0.99\linewidth]{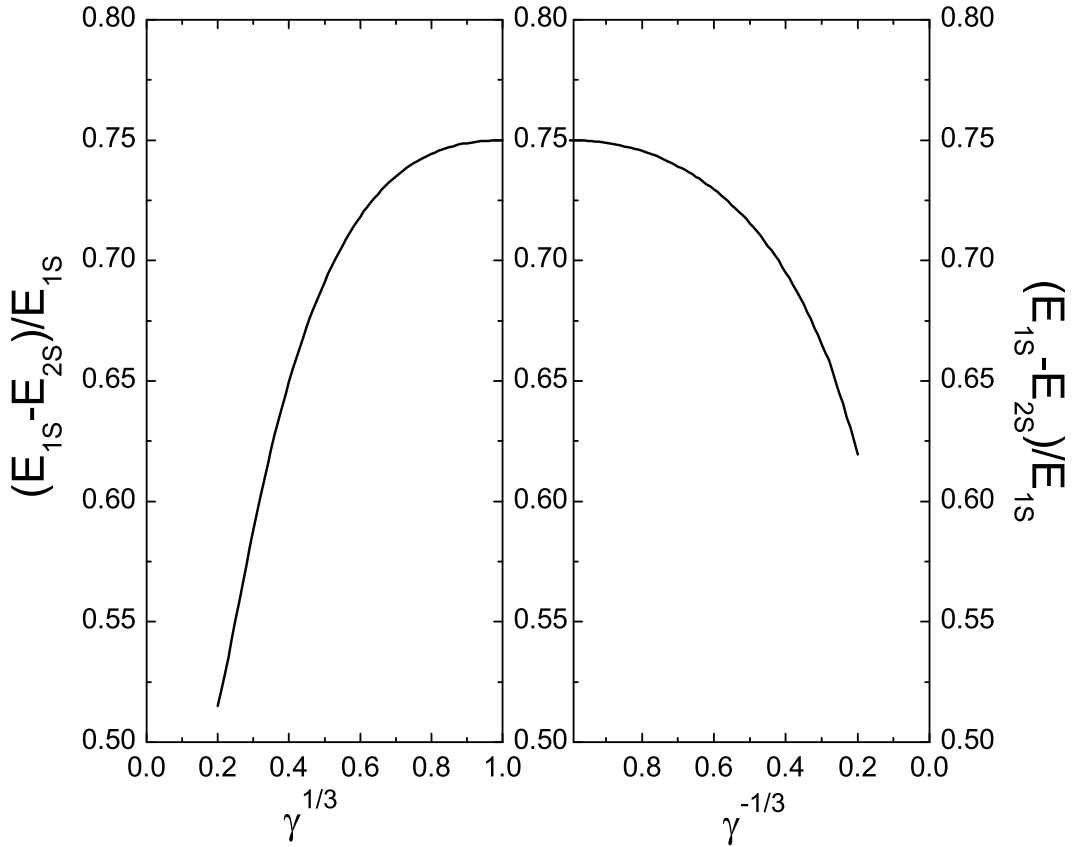}
\vskip-1cm
\caption{
The energy separation between the ground and first excited states in
units of the exciton binding energy vs $\protect\gamma ^{1/3}$.
}
\end{figure}
\begin{table}
\caption{
Exciton binding energies of several lower states calculated by means of
Brillouin-Wigner perturbation method, compared with that taken from
Ref.~\protect\onlinecite{Faul69}.
}\begin{tabular}{l lll lll lll}
        & $\gamma^{1/3} $ &  $1S$  &   $2S$  &   $2P_0$  &  $2P_\pm$
&  $3S$\tablenotemark[1]  &  $3D_0$\tablenotemark[1] & $3P_0$ & $3P_\pm$ \\
\hline
Ref.~\protect\onlinecite{Faul69}   &  0.8          & 1.233   & 0.3151
& 0.3663 & 0.2823      & 0.158  & 0.1375  & 0.1653  &  0.1272\\
This work &               & 1.2327   & 0.3151  & 0.3664 & 0.2823
& 0.1374  & 0.1577  & 0.1652  & 0.1272\\
\hline
Ref.~\protect\onlinecite{Faul69}   &  0.4          & 2.01    & 0.695
& 0.933   & 0.3612      & 0.394  & 0.265   & 0.496   &  0.2100\\
This work &               & 2.011   & 0.6832  & 0.9381 & 0.3615
& 0.2835  & 0.4141  & 0.4959  &  0.2107\\
\end{tabular}
\tablenotetext[1] {The levels classification used in the present work
differs
from that of Ref.~\onlinecite{Faul69}.}
\end{table}

Results of our calculation for several low levels reproduce Faulkner's
calculations~\cite{Faul69} with a good accuracy (see Table~1). As compared
to Faulkner, we calculate a large number of excited states (up to 100 for
each parity and $m$ considered); we calculate the excitonic parameters in
the region of $\gamma \leq 1$ as well as $\gamma \geq 1$, thus covering all
possible values of the anisotropy parameter. Note the difference between
Faulkner's and our designations of $3S$ and $3D_0$ states.~\cite{foot3} When
the states are split off due to perturbation, we always label the states
with larger oscillator strengths at $\gamma \approx 1$ as $S$-state, thus
establishing an order reversed to that among the states with $m\neq 0$,
within our notations (see also discussions in Sec.~\ref{Spherical} and
Fig.\thinspace 5). Thus, at $\gamma <1$ the $3D_0$ level lies lower than
$3S$%
, contrary to the classification by Faulkner.\cite{Faul69} The same
situation holds if we consider the higher excited states.

\subsection{Spherical approximation}

Even in case of small anisotropy $|\epsilon |\ll 1$, the exciton states are
linear combinations of hydrogen states with different $l$. However, for
small $\epsilon $ the admixture of such states becomes rather small, and the
accounting only for the spherically symmetric part of the perturbation
proves to be very useful for understanding the evolution of levels. It is
important that within such a spherical approximation, the anisotropic
exciton problem is exactly soluble.

\begin{figure}
\vskip-1cm
\includegraphics[angle=0,width=0.6\linewidth]{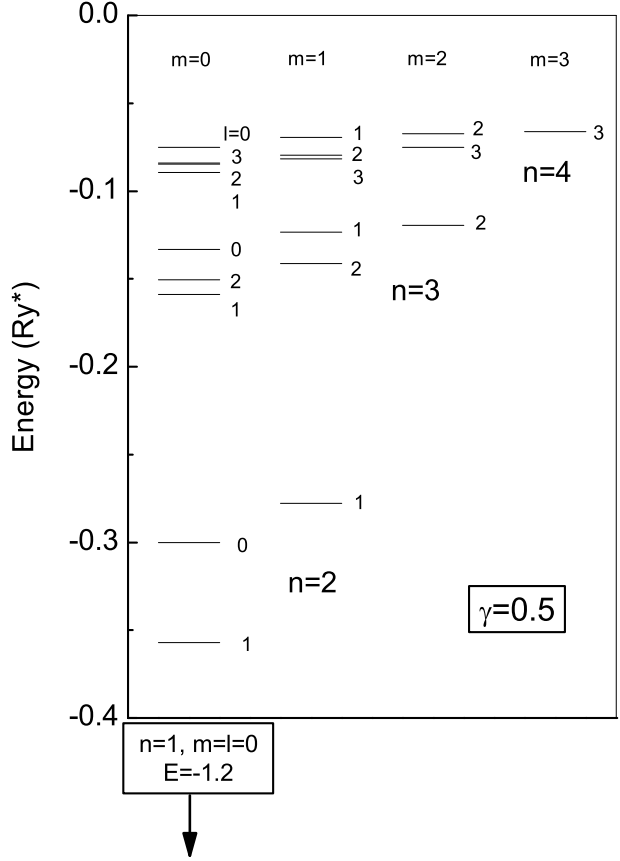}
\vskip-1cm
\caption{
Classification scheme of the energy levels
of the anisotropic exciton
with general quantum number $n\leq4$ in accordance with the spherical
approximation, Eq.~(\protect\ref{E-nlm}).
$\gamma =0.5$.
}
\end{figure}
In this section we consider the approximate solution of the anisotropic
exciton problem in a form $\psi({\bf r})=R(r)Y_{lm}(\theta,\varphi)$, thus
taking into account only diagonal in $l$ parts of the perturbation, Eqs.~(%
\ref{V-ll}),(\ref{B-lm}), and neglecting the perturbation matrix elements
mixing different spherical harmonics.

In order to neglect $l\neq l^{\prime}$ matrix elements, let us replace in
the Schr\"odinger equation, Eq.~(\ref{Schr-p}), the operator $\hat{p}_z^2$
by the operator $\hat{Q}$, defined as
\begin{equation}
\hat{Q} Y_{lm}(\theta,\varphi)=p^2{\cal Q}_{lm}Y_{lm}(\theta,\varphi),
\end{equation}
\begin{equation}
{\cal Q}_{lm}=\int \cos^2\theta|Y_{lm}(\theta,\varphi)|^2 d^3\Omega
\end{equation}
[see also Eq.~(\ref{B-lm})]. Then, after the substitution
\begin{equation}
p\rightarrow \frac{p}{1+\epsilon{\cal Q}_{lm}},\ \ \ p_\nu^2\rightarrow
\frac{p_\nu^2}{1+\epsilon{\cal Q}_{lm}},
\end{equation}
which, in fact, corresponds to a $(l,m)$--dependent mass renormalization, we
arrive at a symmetrical (unperturbed) Schr\"odinger equation with the
solution
\begin{equation}
\label{wf-apprx}\phi_\nu({\bf r})= (1+\epsilon{\cal Q}_{lm})^{-3/2}
\phi^{(0)}_{nlm}\left(\frac{{\bf r}} {1+\epsilon{\cal Q}_{lm}}\right),
\end{equation}
\begin{equation}
\label{E-nlm}E_\nu=-\frac{1}{n^2(1+\epsilon{\cal Q}_{lm})},
\end{equation}
in units of Eq.(\ref{units}) and with the use of dilatation of $z$.

One can easily see from Eq.~(\ref{wf-apprx}) that in this spherical
approximation the perturbation compresses (for $\epsilon<0$) or dilates (for
$\epsilon>0$) the scale of a given hydrogen wave function by the factor $%
1+\epsilon{\cal Q}_{lm}$, which is different for different spherical
harmonics. Note, that the hidden hydrogen-like symmetry is broken within
this spherical approximation, and the binding energies now depend on $l$ and
$m$. However, the spectrum Eq.~(\ref{E-nlm}) still has a hydrogen-like
dependence on the principle quantum number $n$.

In Fig.\,5 we show schematically the energy levels of anisotropic exciton,
calculated via Eq.~(\ref{E-nlm}) for $\gamma=0.5$, $n\leq4$ and all possible
$l$ and $m$. Equation (\ref{E-nlm}) provides a correct qualitative
description of the levels evolution and is in agreement with the result of
calculations presented in Sec.~\ref{energies} in the vicinity of $\gamma =1$
(see Fig.\,6).

\begin{figure}
\vskip-1.5cm
\includegraphics[angle=0,width=0.6\linewidth]{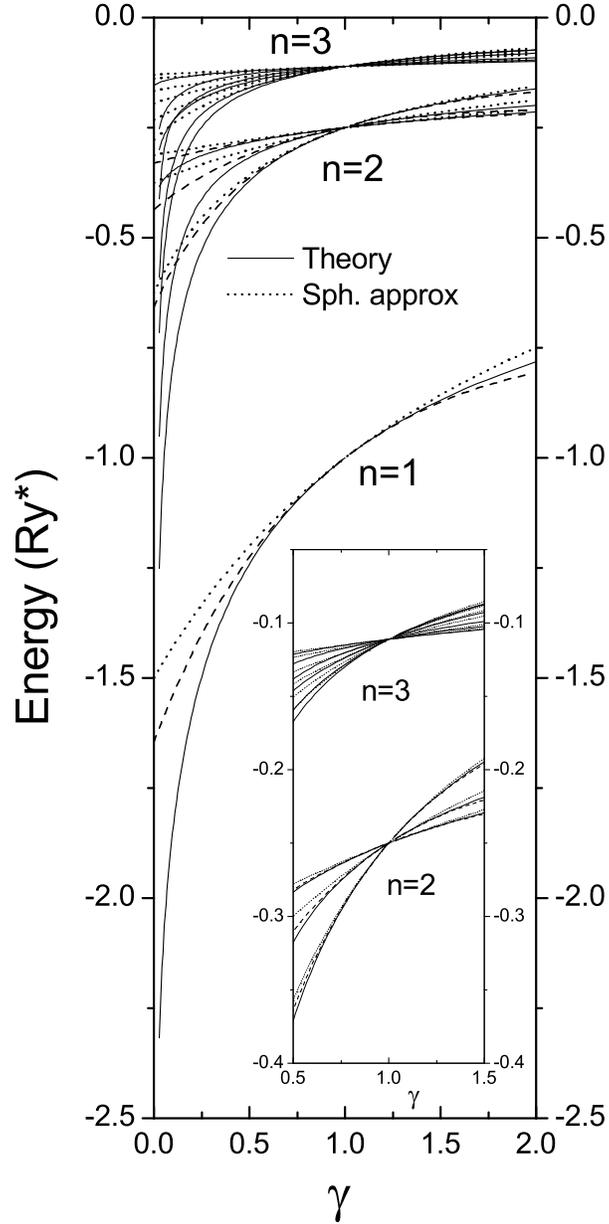}
\vskip-1cm
\caption{
Lower energy levels as functions of
$\protect\gamma^{1/3}$, calculated by means of the perturbation
method (solid curves), within the spherical approximation
(dotted lines), and in the 2nd order perturbation theory
approximation (dashed curves).
}
\end{figure}
The accounting for $l\neq l^{\prime}$ matrix elements (in case of small $%
\epsilon$) yields correct quadratic in $\epsilon$ terms in the energies. The
rational form of matrix elements Eqs. (\ref{V-ll2}) and (\ref{Fnll1}) allows
us to sum up the standard perturbation theory series in the second order.
The calculated in the second order exciton binding energies of several lower
levels are given in Appendix A [see Eq.~(\ref{second})], their dependence on
$\gamma$ is also illustrated in Fig.\,6 (dashed lines).

\subsection{Oscillator strengths}

\label{Spherical}

Within the envelope function approximation, the relative oscillator
strengths of dipole-allowed transitions $f_\nu $ are proportional to $%
|\phi _\nu (0)|^2$ (see, e.g., in Ref.~\onlinecite{Dres57}). Bearing in mind
the expansion of Eq.~(\ref{coordinate}) and the fact that for the
unperturbed states $\phi _\nu (0)\neq 0$ only for $l=m=0$, we get
\begin{equation}
f_\nu \propto |\phi _\nu (0)|^2=\frac{p_\nu ^3}{\pi S_\nu ^2}\left| \sum_n%
{\cal C}_{n,0,0}^\nu \sqrt{n}\right| ^2.
\end{equation}
\begin{figure}
\vskip-1cm
\includegraphics[angle=0,width=0.99\linewidth]{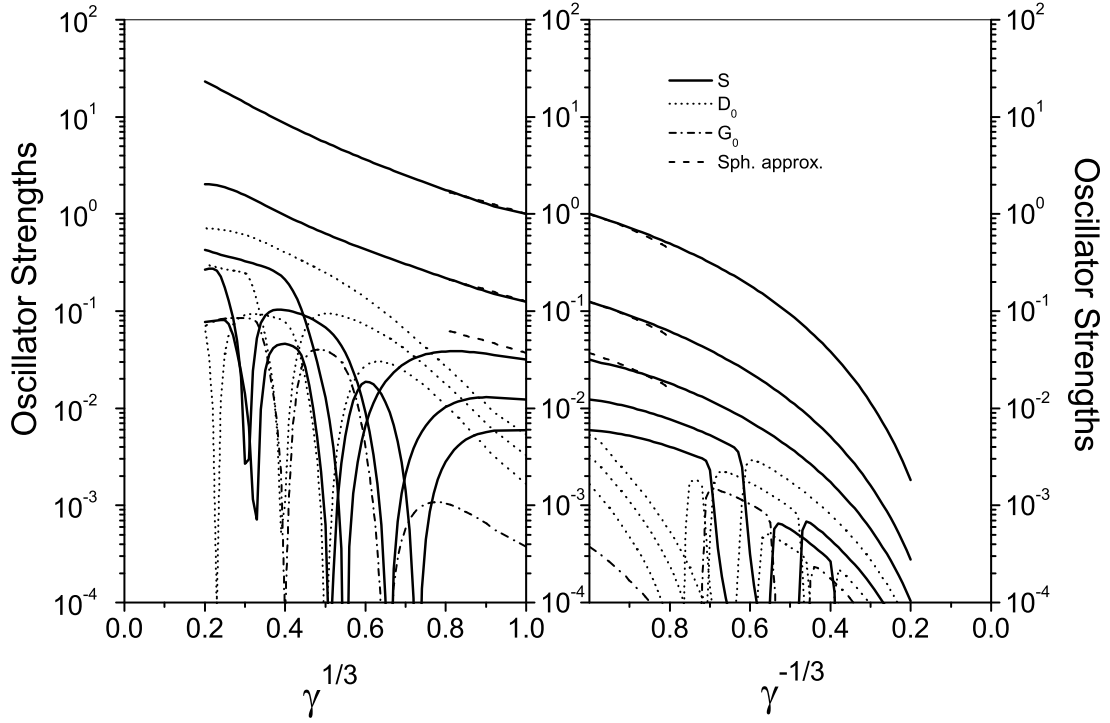}
\vskip-1cm
\caption{
The anisotropic exciton oscillator strengths
of lower $S$, $D_0$ and $G_0$-like  states
as functions of the anisotropy parameter $\protect\gamma^{1/3}$,
calculated numerically and within the spherical approximation
(in units of the ground state oscillator strength
at $\gamma =1$).
Within the spherical approximation, the oscillator strengths
of $D_0$ and $G_0$-like states are vanishing.
}
\end{figure}
Figure~7 shows the calculated numerically oscillator strengths of lower $S$,
$D_0$ and $G_0$-like states as functions of the anisotropy parameter. It is
seen in Fig.\thinspace 7 that the oscillator strengths of all shown states
do not vanish at $\gamma =1$. Originated from the degenerate states of
isotropic 3D exciton, the perturbed states become fixed linear combinations
of the former even when the perturbation tends to zero. It can be explained
as follows. The perturbation of a symmetry lower than the original
Hamiltonian implies the existence of strictly definite combinations of basis
functions for degenerate states when $\gamma =1$, while the symmetry of the
unperturbed Hamiltonian allows an arbitrary choice of these combinations. At
$\gamma \approx 1$ $S$-like state is optically more intensive than $D_0$%
-like state. The picture changes drastically with the increase of
anisotropy. Near $\gamma ^{1/3}=0.8$ the oscillator strength of $3D_0$ state
overcomes that of $3S$ one. For $\gamma ^{1/3}<0.8$ the intensity of the
$3S$
state collapses due to the interaction with $4D_0$ state and then revives
after interaction with higher levels. Moreover, the anisotropy increase
leads to substantial growth of the oscillator strengths of higher excited
states, such as $4S$ and $4D_0$, making them optically significant. Similar
situation takes place if $\gamma >1$ (when a transition from 3D to 1D
exciton occurs). Such a redistribution of the oscillator strengths between
different states is due to multiple unticrossings between energy levels
interacting with each other. This effect can be clearly seen in
Fig.\thinspace 8, where the area of a circle placed on the energy curve is
proportional to the oscillator strength of a given excited state, normalized
to the ground state oscillator strength.

\begin{figure}
\vskip-0.5cm
\includegraphics[angle=0,width=0.7\linewidth]{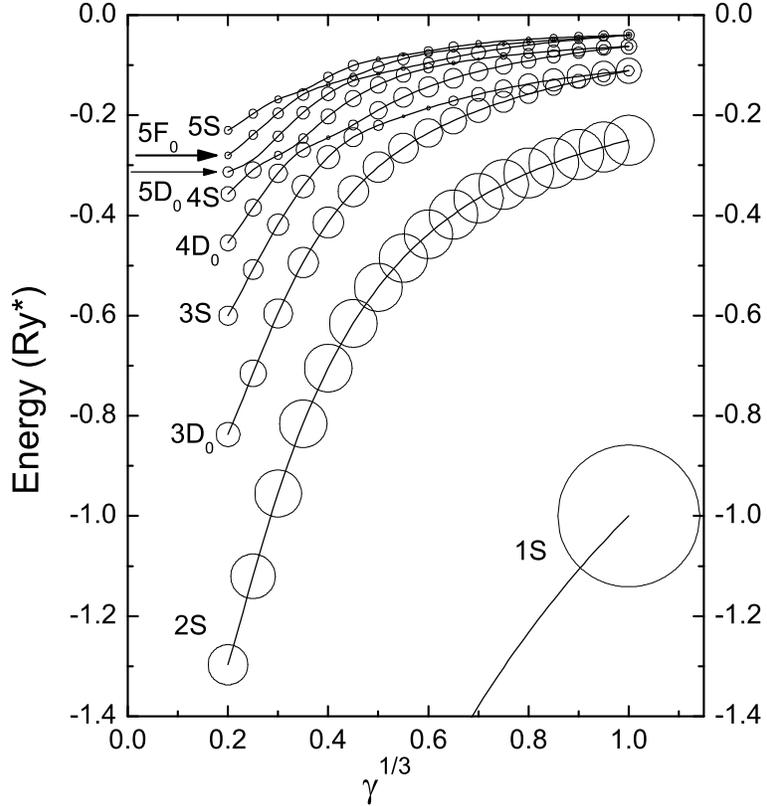}
\vskip-1cm
\caption{
Energies and oscillator strength of excited optically active states
vs $\protect\gamma^{1/3}$ ($\protect\gamma \leq 1$).
The area of a circle is equal to the oscillator
strength, normalized to the ground state oscillator strength, which is
taken constant for all $\gamma $, see a single circle on the bottom
(ground state) curve.
}
\end{figure}
Within the spherical approximation, as it follows from Eq.~(\ref{wf-apprx}),
\begin{equation}
\label{osc-apprx}f_\nu =\left( 1+\frac \epsilon 3\right) ^{-3}f_\nu ^{(0)},
\end{equation}
where $f_\nu ^{(0)}=1/n^3$ are the oscillator strengths of the isotropic
exciton (in units of $f_{1S}^{(0)}$). The oscillator strengths of $1S$, $2S$
and $3S$ states calculated according to Eq.~(\ref{osc-apprx}) are displayed
in Fig.\thinspace 7 by dashed lines. Note that the oscillator strengths of
$%
3S$ state, calculated numerically and within the spherical approximation, do
not coincide at $\gamma =1$, as the spherical approximation does not reflect
correctly the symmetry violation in the vicinity of this point. However, the
sum of the oscillator strengths of $3S$ and $3D_0$ levels is equal to
$1/3^3$.

\section{Conclusions}

The perturbation theory of anisotropic exciton is developed based of the
Fock transformation.
This transformation depends on the exciton energies as
adiabatic parameters and admits a separation of bound and scattering exciton
states. For the bound states the eigenfunctions are expanded into a complete
set of hyperspherical harmonics on a 4D-sphere, creating a representation of
the full symmetry group {\it O(4)} of hydrogen-like system, and the
perturbation matrix elements acquire an explicit algebraic form. This allows
us to analytically perform a partial diagonalization of the Hamiltonian
matrix. It results in a spherical approximation which proves to be very
useful for levels evolution analysis. The eigenvalues and eigenvectors are
found by a numerical diagonalization of the effective Hamiltonian matrix.
The energies and oscillator strengths of anisotropic exciton states are
calculated for all values of the anisotropy parameter $0<\gamma <\infty $
(including both flattened and elongated excitons), except the vicinities of
$\gamma =0$ and $\gamma =\infty $ where the dimensionality of the system
changes, respectively, to $D=2$ and to $D=1$. It is found that with the
increase of the anisotropy a strong redistribution of oscillator strengths
between optically active and formerly inactive states occurs: the
oscillations in optical intensities of higher excited states take place, and
the switching on of formerly weak optical transitions is predicted.

\acknowledgments
The authors are thankful to R. Zimmermann for critical reading of
the manuscript, useful discussions which
helped us to clarify the question of completeness of the basis used in our
perturbation method, and for helpful advices. This work was supported by
Russian Basic Research Foundation, Russian Ministry of Science (program
``Nanostructures''), and INTAS (grant \#96-0398). A. E. B. was supported by
the Dissertation Fellowship from CUNY.

\appendix

\section{ Perturbation matrix. Exciton binding energies in the second order
approximation}

Due to a separation of variables, the matrix element $V_{ss^{\prime}}$,
Eq.~(%
\ref{Vss}), takes the form
\begin{equation}
{V}_{nn^{\prime}}^{ll^{\prime};mm^{\prime}}={\cal J}_{ll^{\prime}}^{mm^{%
\prime}} {\cal I}_{nn^{\prime}}^{ll^{\prime}}\sqrt{nn^{\prime}},
\end{equation}
where
\begin{equation}
{\cal J}_{ll^{\prime}}^{mm^{\prime}}=\delta _{m\,m^{\prime}}{\cal N}_{lm}%
{\cal N}_{l^{\prime}m} \int_{-1}^1 P_l^m(x)P_{l^{\prime}}^m(x)x^2dx,
\end{equation}
\begin{equation}
{\cal N}_{lm}=\left[\frac{(2l+1)}{2}\frac{(l-|m|)!}{(l+|m|)!}\right]^{1/2},
\end{equation}
\begin{equation}
{\cal I}^{ll}_{nn^{\prime}}={\cal D}^\ast_{nl}{\cal D}_{n^{\prime}l}\int^1
_{-1} (1-x^2)^{l+\frac{1}{2}}(1+x)C_{n-l-1}^{l+1}(x)C_{n^{%
\prime}-l-1}^{l+1}(x)dx,
\end{equation}
\begin{equation}
{\cal D}_{nl}=(-2i)^l l! \sqrt{\frac{2n(n-l-1)!}{\pi(n+l)!}}.
\end{equation}
Using the recurrent relations
\begin{equation}
(2l+1)xP_l^m=(l-m+1)P_{l+1}^m+(l+m)P_{l-1}^m,
\end{equation}
\begin{equation}
\label{recurrent1}2(\nu +\alpha )x C^{\nu}_{\alpha }(x)=(\alpha +1)
C_{\alpha +1}^{\nu }(x)+(2\nu +\alpha -1)C_{\alpha -1}^{\nu }(x),
\end{equation}
and normalization property for Legendre ($P^m_l$) and Gegenbauer ($%
C^\nu_\alpha$) polynomials we get
\begin{eqnarray}
{\cal J}_{ll'}^{mm'}=\delta_{mm'}\left\{\frac{1}{2}\left[1+\frac{1-4m^2}
{(2l-1)(2l+3)}\right]\delta _{l\ l'}+
\frac{1}{2l-1}\sqrt{\frac{[l^2-m^2][(l-1)^2-m^2]}{(2l+1)(2l-3)}}
\delta _{l-2\ l'}
\right.
\nonumber\\
+\left.
\frac{1}{2l+3}\sqrt{\frac{[(l+1)^2-m^2][(l+2)^2-m^2]}{(2l+1)(2l+5)}}
\delta _{l+2\ l'}\right\}
\end{eqnarray}
and
\begin{equation}
{\cal I}_{nn^{\prime}}^{ll}=\delta _{nn^{\prime}} +\frac{1}{2}\sqrt{\frac{%
(n-l)(n+l+1)}{n(n+1)}}\delta _{n+1\ n^{\prime}} +\frac{1}{2}\sqrt{\frac{%
(n+l)(n-l-1)}{n(n-1)}}\delta _{n-1\ n^{\prime}}.
\end{equation}
To derive matrix elements ${\cal I}_{n\,n^{\prime}}^{l\,l\pm2}$ we use the
tabulated integral~\cite{Grad62}
\begin{equation}
\label{Int1}\frac{2}{\pi }\int_{-1}^1(1-x)^{-1}(1-x^2)^{\nu -\frac{1}{2}}
C_m^\nu (x) C_n^\nu
(x)dx=\frac{2}{\sqrt{\pi }}\frac{\Gamma(\nu -\frac{1}{2})%
} {\Gamma(\nu )} C_m^\nu(1),\ \ \ m \leq n,
\end{equation}
where
\begin{equation}
C^\nu _m(1)=\frac{(m+2\nu -1)!}{(2\nu -1)! m!},\ \nu \neq0, \ \ \ C^0_m(1)=%
\frac{2}{m},\ m\neq0.
\end{equation}
\begin{equation}
\label{Int2}{\cal I}_{nn^{\prime}}^{l\,l-2}= {\cal D}^\ast_{nl}{\cal D}%
_{n^{\prime}l-2}\int_{-1}^1(1-x^2)
(1+x)^{l-\frac{1}{2}}(1-x)^{l-\frac{3}{2}%
} C_{n-l-1}^{l+1}(x)C_{n^{\prime}-l+1}^{l-1}(x)dx.
\end{equation}
Using Eq.~(\ref{recurrent1}) and the recurrent relations
\begin{equation}
2\nu (1-x^2) C_{\alpha -2}^{\nu +1}(x)= (\alpha +2\nu -1)x C_{\alpha
-1}^\nu(x) -\alpha C_\alpha ^\nu (x),
\end{equation}
\begin{equation}
\alpha C_\alpha ^{\nu -1}(x) =2(\nu -1)\left[xC^\nu _{\alpha -1}(x)-C^\nu
_{\alpha -2}(x)\right],
\end{equation}
we are able to write the integral in Eq.~(\ref{Int2}) in the form of Eq.~(%
\ref{Int1}),
\begin{equation}
{\cal I}_{n\,n^{\prime}}^{l\,l-2}= {\cal D}^\ast_{nl}{\cal
D}_{n^{\prime}l-2}%
\frac{l-1}{l} \int_{-1}^1\!(1+x)^{l-\frac{1}{2}}(1-x)^{l-\frac{3}{2}}
\left(\xi_{nl}\,C^l_{n-l-1}-\eta_{nl}\,C^l_{n-l+1}\right)
\left(\,C^l_{n^{\prime}-l-1}-\,C^l_{n^{\prime}-l+1}\right) dx,
\end{equation}
where
\begin{equation}
\xi_{nl}=\frac{(n+l)(n+l-1)}{4nn^{\prime}}, \ \ \ \ \ \eta_{nl}=\frac{%
(n-l)(n-l+1)}{4nn^{\prime}}.
\end{equation}
After simple transformations we arrive to Eqs.~(\ref{V-ll})--(\ref{Fnll1}).

Note that we are able to apply the standard Reley-Schr\"odinger perturbation
theory to Eq.~(\ref{matrix-equ}) and to calculate analytically the
perturbation theory corrections (to nondegenerate levels) of a given order,
due to the rational form of the perturbation matrix elements
Eqs.~(\ref{V-ll}%
)--(\ref{Fnll1}). For example, for several lower levels the accounting for
the perturbation theory terms up to the second order inclusive leads to
\begin{eqnarray}
\!\!\!\!\!\!E_{1S}=-\frac{1}{\left[1+\frac{1}{6}\epsilon+
\left(\frac{\pi ^2}{45}-\frac{59}{216}\right)\epsilon^2\right]^2},
\ \ \ 
E_{2S}=-\frac{1}{4\left[1+\frac{1}{6}\epsilon+
\left(-\frac{4\pi ^2}{45}+\frac{173}{216}\right)\epsilon^2\right]^2},
\nonumber\\
E_{2P_0}=-\frac{1}{4\left[1+\frac{3}{10}\epsilon+
\left(\frac{4\pi ^2}{35}-\frac{25441}{21000}\right)\epsilon^2\right]^2},
\ \ \ 
E_{2P_\pm}=-\frac{1}{4\left[1+\frac{1}{10}\epsilon+
\left(\frac{8\pi ^2}{105}-\frac{49307}{63000}\right)\epsilon^2\right]^2}.
\label{second}
\end{eqnarray}


\begin{references}
\bibitem[*]{email}  Electronic address: muljarov@gpi.ru

\bibitem{Kitt54}  C. Kittel and A. Mitchell, Phys. Rev. {\bf 96}, 1488
(1954).

\bibitem{Kohn55}  W. Kohn and J. M. Luttinger, Phys. Rev. {\bf 98}, 915
(1955).

\bibitem{Bast88}  G. Bastard, {\it Wave Mechanics Applied to Semiconductor
Heterostructures} (Les Editions de Physique, Les Ulis, France, 1988), p. 26.

\bibitem{Pere90}  M. F. Pereira Jr., I. Galbraith, S. W. Koch, and G.
Duggan, Phys. Rev. B {\bf 42}, 7084 (1990).

\bibitem{Ray93}  Partha Ray and P. K. Basu, Phys. Rev. B {\bf 47}, 15958
(1993).

\bibitem{Hopf61}  J. J. Hopfield and D. G. Thomas, Phys. Rev. {\bf 122}, 35
(1961).

\bibitem{Whee62}  R. G. Wheeler and J. O. Dimmock, Phys. Rev. {\bf 125},
1805 (1962).

\bibitem{Deve69}  J. A. Deverin, Nuovo Cimento B {\bf 63}, 1 (1969).

\bibitem{Sega67}  B. Segal, Phys. Rev. {\bf 163}, 769 (1967).

\bibitem{Faul69}  R. A. Faulkner, Phys. Rev. {\bf 184}, 713 (1969).

\bibitem{Akim67}  O. Akimoto and H. Hasegawa, J. Phys. Soc. Jpn. {\bf 22},
181 (1967).

\bibitem{Kane69}  E. O. Kane, Phys. Rev. {\bf 180}, 852 (1969).

\bibitem{Bald70}  A. Baldereschi and M. G. Diaz, Nuovo Cimento B {\bf 68},
217 (1970).

\bibitem{Zimm71}  R. Zimmermann, Phys. Stat. Sol. (b) {\bf 46}, K111 (1971).

\bibitem{Xia89}  Jian-Bai Xia, Phys. Rev. B {\bf 39}, 5386 (1989).

\bibitem{Depp92}  J. Deppe, M. Balcanski, R. F. Wallis, and K. P. Jain, Sol.
State Commun. {\bf 84}, 67 (1992).

\bibitem{foot1}  As it immediately follows from the form of the anisotropic
exciton Hamiltonian [see Eq.~(\ref{Hamiltonian})], through a substitution of
variables one can make isotropic either the kinetic or the potential energy.

\bibitem{He9091}  X. F. He, Phys. Rev. B {\bf 42}, 11751 (1990); {\bf 43},
2063 (1991).

\bibitem{Tang97}  Ch. Tanguy, P. Lefebvre, H. Mathieu, and R. J. Elliot,
Phys. Stat. Sol. (a) {\bf 164}, 159 (1997).

\bibitem{Pere95}  M. F. Pereira Jr., Phys. Rev. B {\bf 52}, 1978 (1995).

\bibitem{our}  A short description of our method see in: E. A. Muljarov, A.
L. Yablonskii, S. G. Tikhodeev A. E. Bulatov and Joseph L. Birman, Phys.
Rev. B, to be published (1999).

\bibitem{Fock35}  V. A. Fock, Zh. Physik {\bf 98}, 145 (1935).

\bibitem{Grad62}  I. S. Gradshtein and I. M. Ryzhik, {\em Tables of
Integrals, Series, and Products} (Academic, New York, 1980).

\bibitem{Band66}  M. Bander and C. Itzykson, Rev. Mod. Phys. {\bf 38}, 330
(1966); {\bf 38}, 346 (1966).

\bibitem{foot11}  It can be shown that $\int \cos \alpha |\Psi
_{nlm}^{(0)}|^2d^4\Omega =0$.

\bibitem{Podo29}  B. Podolansky and L. Pauling, Phys. Rev. {\bf 34}, 109
(1929).

\bibitem{Land76}  L. D. Landau and E. M. Lifchitz, {\em Quantum Mechanics.
Nonrelativistic Theory} (Pergamon Press, New York, 1976).

\bibitem{foot00}  For instance, in the coordinate representation the $S$%
-type basic functions are proportional to $\exp (-p_\nu r)L_n^1(2p_\nu r)$
instead of $\exp (-r/n)L_n^1(2r/n)$, thus forming a complete set for
spherically symmetric functions [See also Eq.~(\ref{coordinate})].

\bibitem{foot2}  It is well known that in exactly 1D case the ground state
exciton energy is infinite (logarithmically diverges). See, e.g., in Ref.~%
\onlinecite{Land76}.

\bibitem{foot3}  The standard quantum numbers $n,l$ and hydrogen-like
notations can be used in case of the anisotropic exciton only approximately.

\bibitem{Dres57}  G. Dresselhaus, Phys. Rev. {\bf 106}, 76 (1957).
\end{references}
\end{document}